# Using Fast-Reading Current Integrator for Advanced Ion Beam Diagnostics Across Continuous and Pulsed Modes

I-Chun Cho[1], Chien-Hsu Chen[2,*], Huan Niu[2], Cheng-Ya Pan[1], Chun-Hui Hsing[1] and Tung-Yuan Hsiao[2]

[1)]Radiation Research Core Laboratory, Institute for Radiological Research, Chang Gung University / Chang Gung Memorial Hospital, Linkou Branch, Taoyuan, Taiwan

[2]Accelerator Laboratory, Nuclear Science and Technology Development Center, National Tsing Hua University, Hsinchu, Taiwan

[*]Author to whom correspondence should be addressed: achchen@mx.nthu.edu.tw

**Abstract**

A fast-reading current integrator is developed for high time-resolution and low-noise ion beam diagnostics under both continuous-wave (CW) and pulsed operating conditions. The system combines a low-leakage transimpedance front-end with a hybrid digitization architecture based on charge-balancing and voltage-to-frequency (V–F) conversion. The input current is converted into a pulse stream corresponding to discrete charge quanta, enabling event-driven measurement with temporal resolution down to 0.5 ms while preserving a wide dynamic range and high linearity.

The system further enables real-time pulse selection for instantaneous dose-rate estimation and reconstruction of time-dependent beam structures. A deterministic beam-interrupt signal is generated within <1 µs upon reaching a predefined charge threshold, enabling fast feedback control. Additional functionalities, including threshold- and slope-based gating as well as phase-sensitive detection, enhance performance under noisy or modulated beam conditions. Calibration with precision current sources and beamline detectors demonstrates stable operation with excellent linearity and timing fidelity. The proposed system provides a compact and flexible platform for next-generation ion beam diagnostics requiring fast response, large dynamic range, and time-resolved measurement.

## 1. Introduction

Accurate measurement of ion beam current with both high temporal resolution and low noise is essential in accelerator diagnostics, ion implantation, ion beam analysis (IBA/IBIC), and beamline instrumentation. Conventional current integrators[1-10] based on capacitor charge–discharge mechanisms achieved high precision but are inherently limited in time-resolved applications due to latency associated with analog readout, dielectric absorption, and drift.

The development of charge-balancing (recycling) integrators[6] introduced a paradigm shift by enabling continuous, dead-time-free charge measurement through periodic removal of fixed charge quanta. In parallel, voltage-to-frequency (V–F) conversion techniques allowed analog current signals to be encoded into pulse rates, facilitating robust digital counting with reduced susceptibility to drift. Despite these advances, most existing systems remain optimized for accumulated charge measurement rather than real-time beam diagnostics. Modern accelerator applications—including pulsed beam delivery, fast scanning, and high-dose-rate irradiation—require not only accurate integration but also rapid extraction of temporal beam characteristics and deterministic control.

In this work, a fast-reading current integrator is presented that unifies charge integration, time-resolved diagnostics, and real-time control. By encoding beam current into a high-rate pulse stream representing discrete charge quanta, the system enables simultaneous measurement of accumulated dose and instantaneous beam behavior. The overall architecture, illustrated in Fig. 1, integrates a low-leakage analog front-end, a hybrid charge-balancing/V–F digitization core, and a hardware-based pulse counting backend, forming a coherent event-driven measurement platform.

The proposed system builds upon the well-established charge-balancing integrator concept originally formalized by Bernard Gottschalk[6], in which a fixed quantum of charge is periodically removed from the integrator node while generating a corresponding output pulse. This approach provides inherent linearity, wide dynamic range, and the absence of intrinsic dead time, and has formed the basis of many subsequent digital current integrators. However, such early implementations were primarily designed for accurate charge accumulation, with limited emphasis on time-resolved measurement or real-time system response.

More recent developments, such as the multi-channel ion beam monitoring system reported by K. Suresh[9], have extended this concept toward practical accelerator instrumentation by integrating charge-balancing V–F conversion with microcontroller-based readout, enabling multi-channel operation and improved measurement stability. These systems demonstrate excellent linearity and long-term stability over a wide current range, but remain largely oriented toward averaged current and fluence measurement, with bandwidth and temporal resolution constrained by system architecture and data acquisition schemes.

In contrast, the present work introduces an event-driven architecture that explicitly targets time-resolved beam diagnostics and deterministic control. By treating each charge-balancing pulse as a discrete event and leveraging asynchronous hardware counting, the system enables sub-millisecond temporal resolution without sacrificing dynamic range or linearity. Furthermore, the integration of real-time pulse processing, flexible gating, and hardware-level beam-interrupt generation (<1 µs latency) extends the functionality beyond conventional integrators, which typically rely on software-mediated control with significantly higher latency.

Another key distinction lies in the unification of measurement and control within a single architecture. While prior systems focus either on precise charge integration or on system-level data acquisition, the proposed design bridges these domains by enabling simultaneous charge accumulation, instantaneous dose-rate estimation, and real-time feedback. This capability is particularly relevant for modern beam delivery scenarios involving fast modulation, pulsed operation, and emerging high-dose-rate regimes, where both temporal fidelity and deterministic response are critical.

Overall, the proposed integrator preserves the fundamental advantages of classical charge-balancing techniques while extending their applicability toward high-speed, time-resolved, and feedback-driven beam diagnostics, representing a functional evolution rather than a purely incremental improvement over existing designs.

## 2. Methods

2.1 System Architecture

The system adopts an event-driven measurement paradigm in which the accumulated charge is determined by counting discrete pulses, each corresponding to a fixed charge quantum. This approach follows the principle of charge-balancing integration, where the total charge is expressed as:

$$Q = N \cdot \Delta Q_{ref}$$

where N is the number of pulses and $\Delta Q_{ref}$ is the reference charge per event.

As depicted in Fig. 1, the analog front-end consists of a low-input-bias transimpedance amplifier (TIA)[6] that converts the ion beam current into a voltage signal while minimizing leakage and noise. Guarding and air-wiring techniques are employed to support measurements down to the pA regime. The integrator output feeds a high-speed logic-based oscillator, which generates a pulse whenever a predefined threshold is reached. At the same instant, a fixed charge packet is injected back into the summing node, maintaining charge balance without interrupting the input current. This mechanism ensures continuous operation without intrinsic dead time and preserves linearity over a wide dynamic range. The pulse generation stage is implemented using fast CMOS logic (74HC/74AHC families), as illustrated in Fig. 1, to reduce propagation delay and timing jitter. The resulting pulse stream, representing discrete charge quanta, is directly fed into a hardware timer within a microcontroller (Arm® Cortex®-M Series). Pulse counting is performed asynchronously (Fig. 2), ensuring that measurement accuracy is independent of CPU latency. Counter overflow is handled through multi-word extension, enabling effectively unlimited counting range. Data acquisition is decoupled from pulse counting by periodic sampling of accumulated counts, allowing continuous measurement even at high pulse rates. To guarantee deterministic communication timing, fixed-length binary encoding is adopted.

2.2 Real-Time Measurement Concept

A defining feature of the proposed system is its ability to perform time-resolved analysis on the pulse stream. Within a configurable acquisition window (typically 0.5 ms), the system evaluates incoming pulses to estimate instantaneous beam current and dose rate. Because each pulse represents a fixed charge quantum, the current can be directly inferred from pulse density within the window.

As illustrated in Fig. 3, the pulse density within each time window directly reflects the beam intensity, enabling:

- Reconstruction of pulsed beam structures
- Detection of beam fluctuations and instability
- Real-time dose-rate estimation

Simultaneously, the accumulated charge is continuously compared against predefined thresholds. When the preset dose is reached, a hardware-triggered beam-interrupt signal is issued with a latency below 1 µs (See Fig. 2). This deterministic response enables precise beam control and enhances operational safety.

2.3 Deterministic Beam Control

In addition to passive measurement, the system enables active beam control. As shown in Fig. 2, the accumulated pulse count is continuously compared to a predefined threshold. When the target dose is reached, a hardware-triggered beam-interrupt signal is generated.

The latency between threshold crossing and output trigger is below 1 µs, determined solely by comparator response and logic propagation delay. This hardware-based implementation ensures deterministic behavior independent of software execution.

2.4 Gating and Phase-Sensitive Detection

To extend functionality under complex beam conditions, flexible gating mechanisms are implemented (Fig. 2 and 3). The system supports:

- Threshold-based gating (amplitude selection)
- Slope-based gating (transient detection)
- External synchronization (phase locking)

Furthermore, a lock-in-style detection scheme is realized digitally by synchronizing pulse accumulation with an external reference signal. By performing phase-sensitive counting, weak modulated signals can be extracted from noisy backgrounds. The high pulse rate provided by the V–F stage enhances phase resolution and reduces timing uncertainty.

2.5 Calibration and Validation

Calibration is performed using precision current sources over a wide dynamic range from pA to µA. As shown in Fig. 4, the pulse rate exhibits a linear relationship with input current, confirming preservation of the fixed charge quantum.

Independent validation using beamline detectors (e.g., Faraday cups and ionization chambers) demonstrates strong agreement, confirming both accuracy and temporal fidelity.

## 3. Results

### 3.1 Bench Tests and Functional Demonstrations

3.1.1 Linearity and Dynamic Range

The linearity and dynamic range of the system were evaluated using a precision DC current source (Keithley 6221) over approximately twelve decades of input current. In charge-balancing mode, the integrator demonstrates a continuous and stable operating range extending from ~100 pA to the μA regime, without the need for manual range switching.

The measured pulse rate exhibits a strictly linear dependence on the input current, consistent with the relation:

$I = f \cdot \Delta Q_{ref}$

As illustrated in Fig. 4, the linearity deviation remains within ±0.1% across the calibrated range. Unlike conventional systems, no range switching is required.

At higher currents, the system maintains linearity until approaching the maximum pulse processing rate, beyond which saturation manifests as a compression of the pulse frequency. However, due to the absence of intrinsic dead time in the charge-balancing architecture, transient overload conditions do not lead to charge loss, but rather to a temporary increase in pulse density. This behavior ensures robustness in applications involving rapidly varying beam intensities.

3.1.2 Temporal Resolution and Time-Structured Beam Reconstruction

The system's temporal response is governed by its event-driven pulse generation and configurable acquisition window. With a 0.5 ms sampling interval, it resolves beam current variations on the millisecond timescale.

As shown in Fig. 5, experimental tests under pulsed beam conditions confirm that the integrator accurately reconstructs time structures from 1 ms to 199 ms. The reconstructed beam profiles align closely with external timing references ($R^2=0.99982$), validating the pulse-counting approach's ability to preserve temporal information without distortion from analog filtering or averaging. Unlike conventional analog integrators—limited by RC time constants and readout latency—this system employs a novel paradigm: beam current is encoded as a time-resolved pulse stream, providing direct, undistorted access to instantaneous beam behavior.

Fig. 6 illustrates the temporal structure of FLASH proton beam pulses from the CGMH SHI 230 MeV proton therapy machine[11], as measured by the fast-reading current integrator. The plot depicts beam current versus time for nominal pulse durations of ~1 ms to 199 ms, with delivered lengths closely matching requests. Each pulse shows a slight downward current trend, and the longest pulse (192.50 ms, requested as 199 ms) is highlighted with a black bracket.

3.3 Real-Time Dose Rate Estimation

Instantaneous dose rate is estimated by counting pulses within short acquisition windows and converting the result into charge per unit time. Since each pulse corresponds to a fixed charge quantum, this method provides a direct and calibration-stable measurement of dose rate.

Fig. 3 Real-time current measurement (differential counted charge). The upper panel shows the automatically captured pulse current when the change in counted charge exceeds a predefined threshold. The system simultaneously calculates the total pulse charge (dose), pulse height, and average pulse current to verify the validity and correctness of the detected output pulse. The lower panel displays the long-term real-time current waveform for context.

The system achieves stable dose-rate estimation across a wide dynamic range, from low-current continuous beams to high-intensity pulsed regimes.

The absence of analog integration drift, combined with hardware-based counting that removes fixed charge quanta, ensures that the measurement is largely insensitive to long-term offset variations.

Furthermore, the combination of high pulse rates and short acquisition windows enables the detection of rapid fluctuations in beam intensity, which are typically averaged out in conventional measurement systems. This capability is particularly valuable for modern beam delivery schemes involving fast scanning or modulation.

3.4 Beam-Interrupt Latency and Deterministic Control

A key feature of the proposed system is its ability to generate a beam-interrupt signal with deterministic latency (<1 µs) upon reaching a predefined charge threshold. This delay encompasses comparator response, logic propagation, and output driver time—and remains independent of software execution.

Fig. 2 depicts the system architecture, including data acquisition, readout electronics, and digital signal output for the fast-reading current integrator. The diagram highlights the Arm® Cortex®-M series microcontroller-based system, which simultaneously acquires pulses from the integrator (via external counter) and timing signals. Count values are periodically read using hardware timers and interrupts, buffered in RAM, and transferred as high-speed binary blocks to a PC via USB 2.0 USART for real-time or offline analysis.

This performance markedly outperforms traditional systems reliant on software polling or serial communication, which typically exhibit millisecond-range response times. The fast interrupt capability enables precise dose control and enhances safety, especially in applications demanding strict dose limits or rapid beam shutdown.

3.6 Long-Term Stability and Cross-Validation

Long-term stability tests over 8–24 hours revealed a measured charge drift below $10^{-4}$, confirming excellent performance comparable to established digital integrators. Cross-validation against independent beamline detectors (e.g., ionization chambers and Faraday cups) showed strong agreement in both instantaneous current and accumulated charge across diverse conditions, underscoring the reliability of the event-based integration approach.

Linearity Verification

Fig. 7 verifies the linearity of the fast-reading current integrator for FLASH proton beam dosimetry. It compares doses measured by the PPC05 ionization chamber (y-axis) versus the integrator (x-axis) for pulses from the 230 MeV SHI proton therapy machine at CGMH, demonstrating excellent linearity across the tested range.

Fig. 8 examines dose linearity with pulse width for FLASH proton beams. It plots PPC05 ionization chamber dose versus set pulse width (1–200 ms) at fixed 30 nA beam current from the same machine, confirming strong proportionality between duration and dose.

Pulse Duration and Charge Relationships

Fig. 9 illustrates the relationship between measured pulse duration and integrated charge (ΔQ counts) under constant beam current for 230 MeV FLASH proton beams at CGMH. Even for identical requested pulse widths, both actual durations and total charge vary noticeably during continuous delivery.

Fig. 10 extends this analysis, plotting measured integrated charge (ΔQ counts) versus actual pulse duration for the same beams at varying currents. It highlights persistent variations in delivered duration and charge (dose), despite uniform requested widths, under dynamic conditions.

## 4. Discussion

Experimental results demonstrate that the proposed system achieves a rare combination of performance metrics—wide dynamic range, high linearity, sub-millisecond temporal resolution, and <1 µs response time—unmatched by conventional current integrators. Crucially, these stem not from incremental circuit tweaks, but from an event-driven charge quantization paradigm.

Traditional analog integrators accumulate voltage continuously before periodic readout, forcing trade-offs between noise, bandwidth, and resolution: long time constants suppress noise but distort fast beam fluctuations (e.g., Ortec 439[3]), while short ones limit real-time response due to analog settling and acquisition latency.

In contrast, this system converts input current to a pulse stream, where each event quantifies a fixed charge packet—adjustable via transistor discharge capacitor to minimize integration time constants and suppress distortion. This decouples charge measurement from analog bandwidth limits, preserving temporal information in pulse density and timing for distortion-free reconstruction of structured beams. Such capability is vital for pulsed or modulated beams requiring transient resolution.

The approach also eliminates intrinsic dead time. Unlike reset-type integrators with periodic discharge gaps, charge-balancing maintains continuous integration by subtracting discrete packets, ensuring no charge loss even at high rates—essential for accuracy under intense or varying beams.

The <1 µs beam-interrupt latency underscores another edge: hardware-triggered interrupts deliver deterministic response independent of processing load, far surpassing software-dependent systems' millisecond delays. This enables precise dose control and safe shutdown in critical applications like ion implantation or medical delivery.

Though not demonstrated here, external counter interrupts can drive GPIO for gating and lock-in detection, extending functionality to isolate beam components or extract weak signals from noise via time/phase-selective accumulation. This leverages the pulse stream's discreteness and high timing fidelity for advanced diagnostics.

Limitations include: (1) maximum current capped by pulse processing rate, risking saturation/nonlinearity at extremes (mitigable via quantum adjustment or parallel channels, trading range for resolution); (2) quantization noise in low-current sparse regimes, addressable by adaptive windows, statistical filtering, or coincidence rejection; (3) microcontroller (Cortex-M with DSP instructions) constraints, improvable with higher-frequency processors, multi-channel scaling, or FPGA for ultra-low latency and real-time computation—ideal for large detectors or high-flux lines.

Ultimately, this system represents a unified framework integrating charge integration, temporal diagnostics, and real-time control, aligning with demands for high-speed precision in accelerators and radiation apps, including FLASH regimes.

To our knowledge, no prior ion beam integrator combines sub-ms timing, hardware event digitization, and <1 µs deterministic control in one architecture. While Gottschalk's charge-balancing and Suresh's multi-channel extensions excel in linearity/range, they prioritize accumulated charge with software-limited observability.

This work shifts charge quantization to a real-time event stream, enabling direct instantaneous access sans analog averaging. Coupled with async hardware counting and full-hardware thresholding, it yields sub-1 µs interrupts free of firmware overhead. Unifying measurement, analysis, and control in one layer unlocks real-time dose estimation, pulse gating, and phase detection—transforming passive accumulation into active, time-resolved diagnostics/control, poised to impact fast-modulated, pulsed, and ultra-high dose-rate applications.

## 5. Conclusion

We developed and experimentally validated a fast-reading current integrator based on event-driven charge quantization for advanced ion beam diagnostics. By integrating charge-balancing with high-rate pulse encoding, the system converts beam current directly into a discrete pulse stream, enabling simultaneous charge measurement and time-resolved analysis.

The architecture eliminates intrinsic dead time, delivering excellent linearity ($\leq 0.1\%$), wide dynamic range (pA to µA), and long-term stability ($\sim 10^{-4}$). Unlike conventional integrators, it unifies precision metrology, real-time diagnostics, and deterministic control—reconstructing beam temporal structures, estimating instantaneous dose rates, and generating <1 µs beam-interrupt signals for fast feedback and safety-critical applications.

Flexible gating and phase-sensitive detection further boost performance in noisy environments. Overall, this compact system offers a versatile solution for high-speed, high-precision accelerator instrumentation.

Acknowledge

This work was supported by the Ministry of Science and Technology, Taiwan under grants no. NSCT-114-NU-E-007-003-NU and NSCT-115-NU-E-007-003-NU.

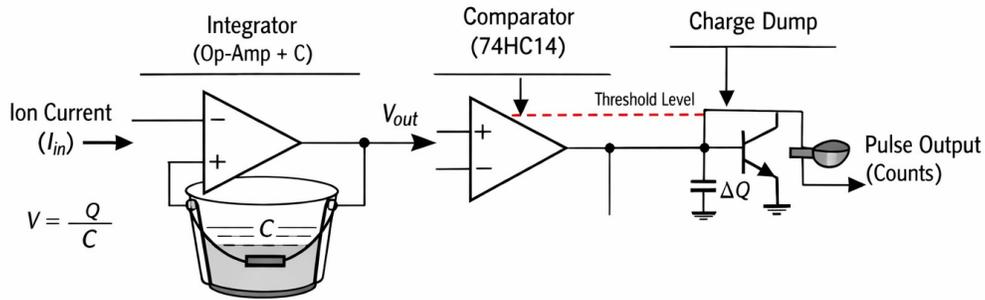

Figure 1

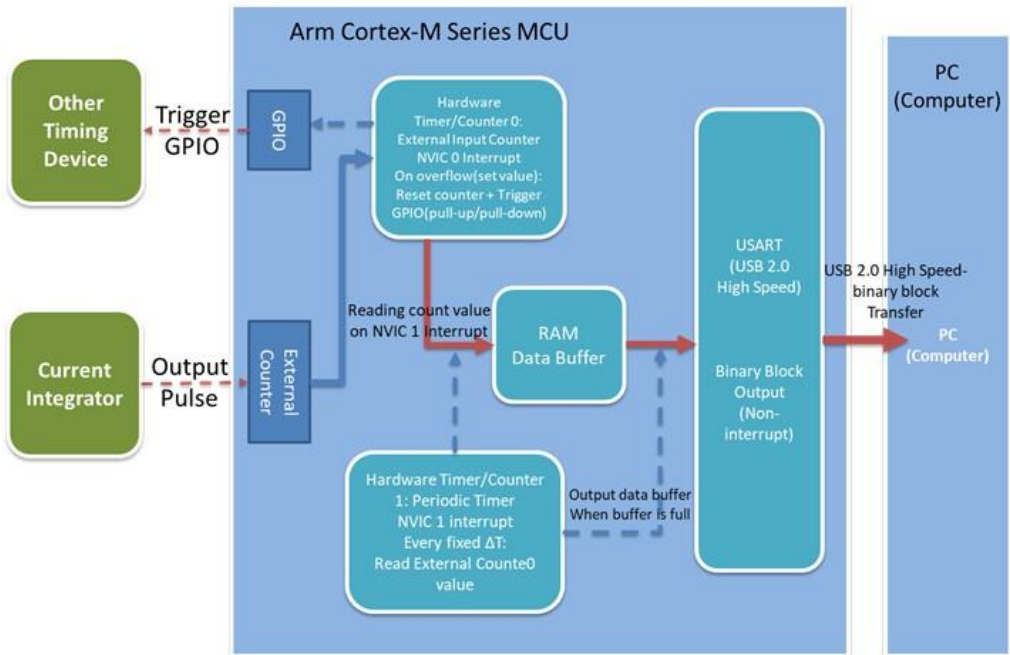

Figure 2

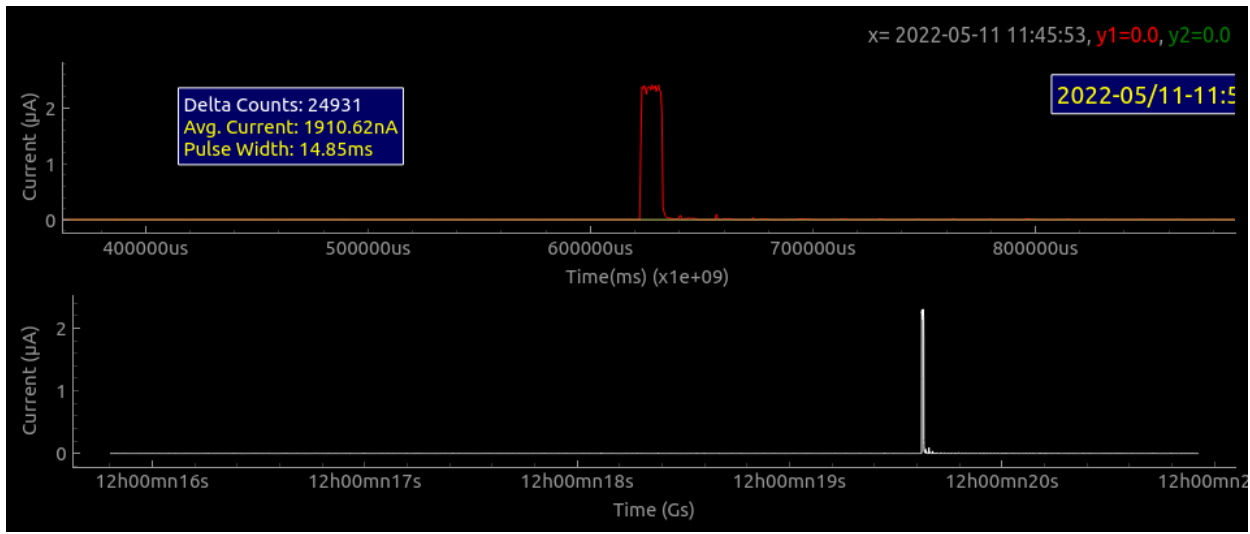

Figure 3.

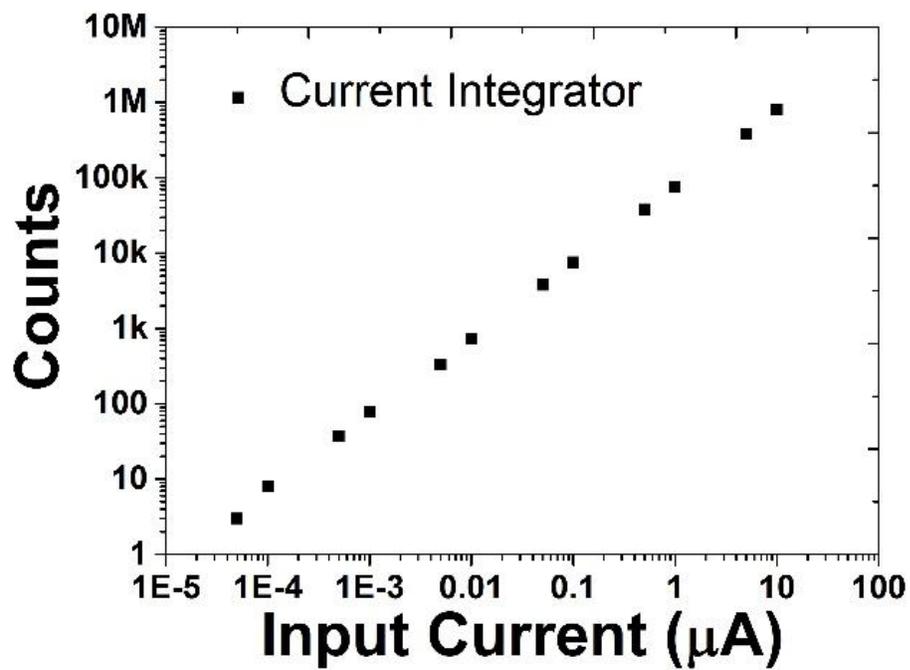

Figure 4.

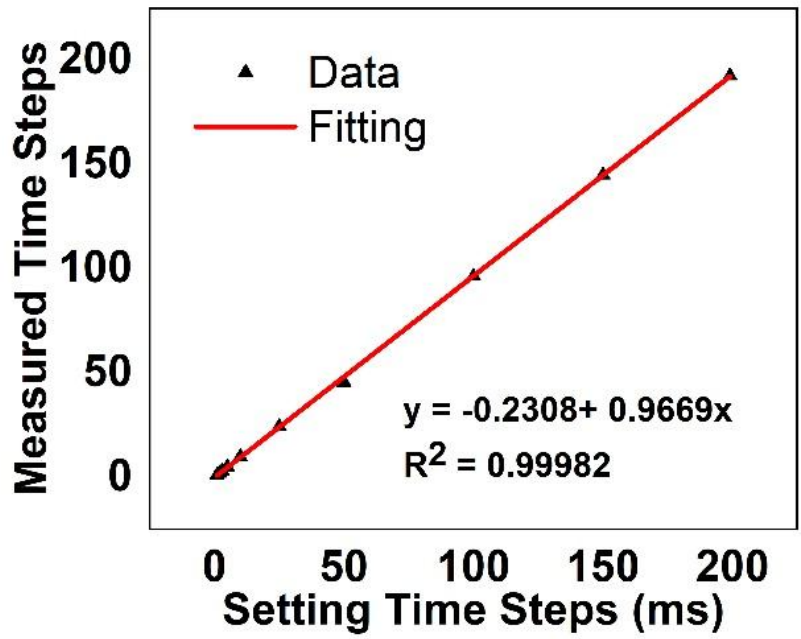

Figure 5.

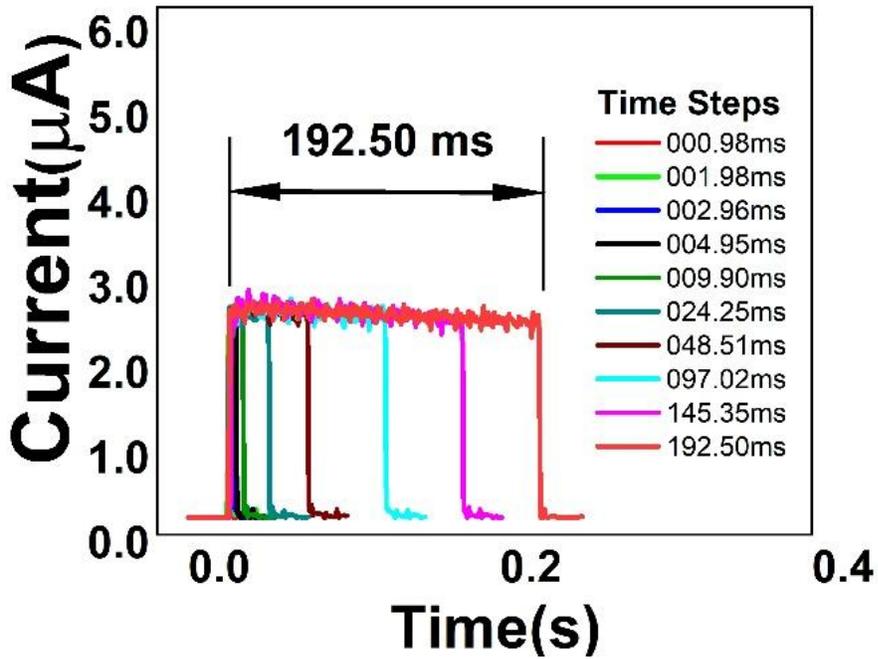

Figure 6.

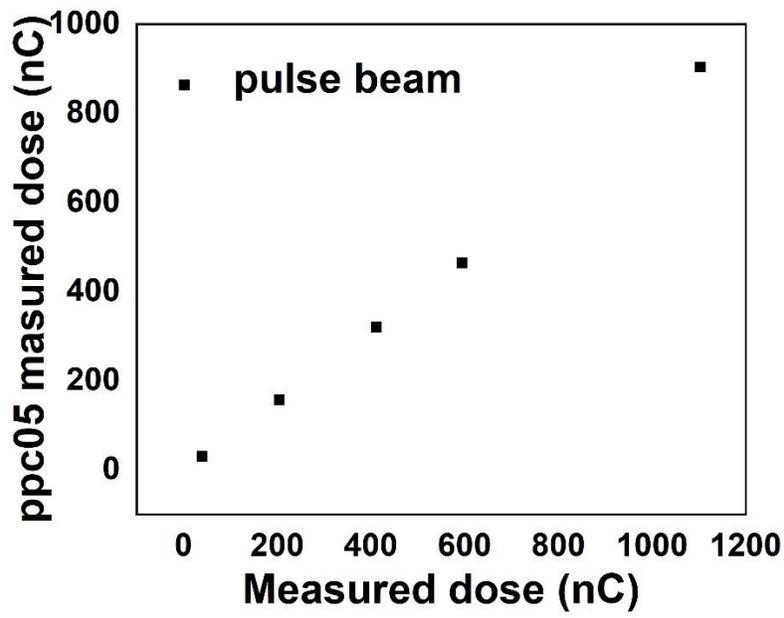

Figure 7.

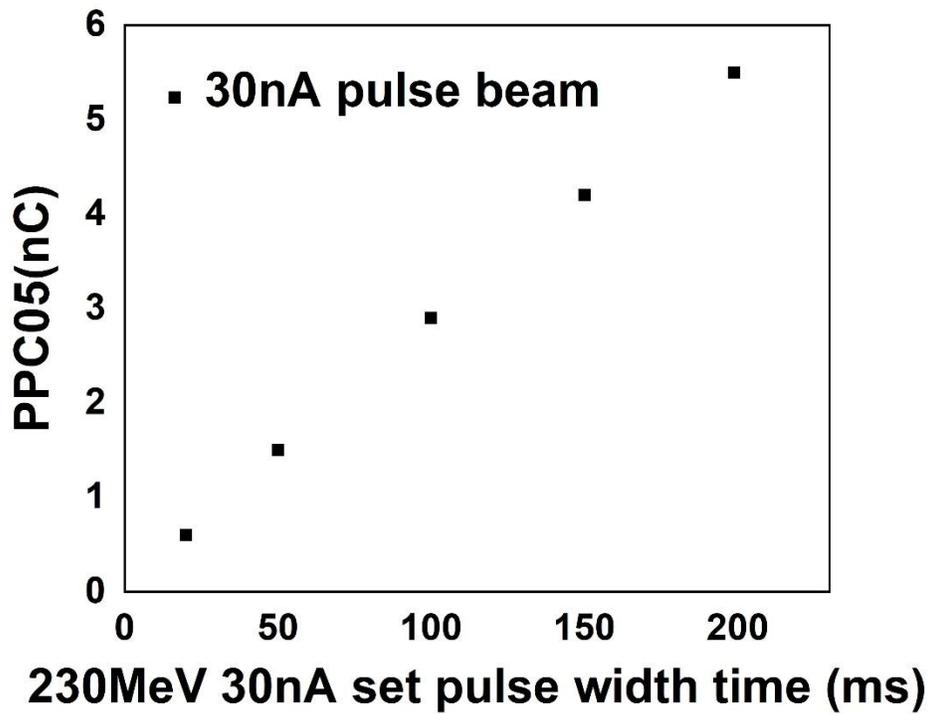

Figure 8.

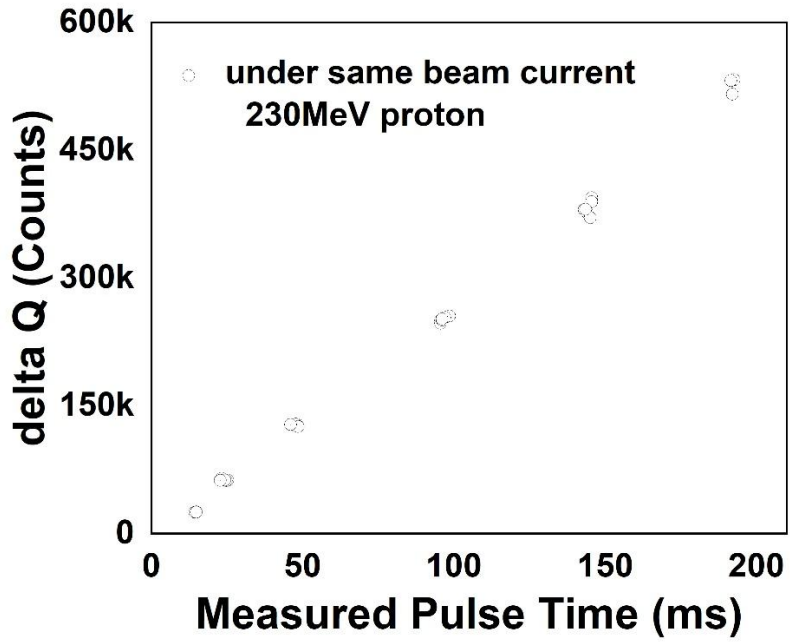

Figure 9.

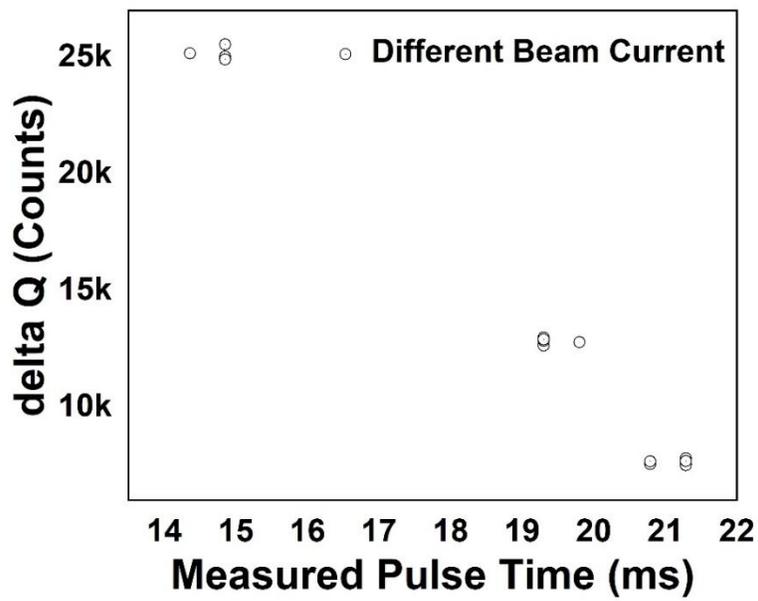

Figure 10.

Figure Caption:

Figure 1. Overall system architecture of the fast-reading current integrator. The system integrates a low-leakage transimpedance amplifier (TIA), a charge-balancing/V–F conversion core, and a hardware-based pulse counting backend. The event-driven architecture enables simultaneous charge accumulation, time-resolved measurement, and real-time beam control.

Figure 2. System architecture of the data acquisition and readout electronics for the current integrator. The diagram illustrates the Arm® Cortex®-M series microcontroller-based system that simultaneously acquires pulses from the current integrator (via external counter) and timing signals from the current integrator. Count values are periodically read using hardware timers and interrupts, buffered in RAM, and transferred as high-speed binary blocks to a PC via USB 2.0 USART for real-time or offline analysis.

Figure 3. Real-time current measurement (differential counted charge). The upper panel shows the automatically captured pulse current when the change in counted charge exceeds a predefined threshold. The system simultaneously calculates the total pulse charge (dose), pulse height, and average pulse current to verify the validity and correctness of the detected output pulse. The lower panel displays the long-term real-time current waveform for context.

Figure 4. Calibration of the current integrator using the Keithley 6221 DC/AC current source meter. The log-log plot shows the linear relationship between input current (µA) and output counts.

Figure 5. Linearity of FLASH proton beam pulse width delivered by the SHI proton therapy system at LK-CGMH. The plot compares the nominal (setting) pulse duration versus the actual measured pulse duration. A linear fit yields a slope of 0.9669 with an excellent correlation coefficient ($R^2$ = 0.99982), demonstrating highly accurate and reliable pulse timing control across the range of 1 ms to 199 ms.

Figure 6. Temporal structure of FLASH proton beam pulses from the CGMH SHI 230 MeV proton therapy machine, measured using a fast-reading current integrator. The plot shows the measured beam current versus time for nominal pulse durations ranging from approximately 1 ms to 199 ms. The actual delivered pulse lengths closely match the requested timings. A slight decreasing trend in current is observed during each pulse. The longest pulse (192.50 ms, requested as 199 ms) is highlighted with a black bracket.

Figure 7. Linearity verification of the fast-reading current integrator for FLASH proton beam dosimetry. The plot compares the dose measured by the PPC05 ionization chamber (y-axis) against the dose measured by the fast-reading current integrator (x-axis) for pulse beams delivered by the 230 MeV SHI proton therapy machine at CGMH. The results demonstrate good linearity across the tested dose range.

Figure 8. Linearity of delivered dose with pulse width for FLASH proton beams. The plot shows the dose measured by the PPC05 ionization chamber as a function of the set pulse width (1 ms to 200 ms) for a fixed 30 nA pulse beam delivered by the 230 MeV SHI proton therapy machine at CGMH. The results demonstrate excellent linearity between pulse duration and measured dose.

Figure 9. Relationship between measured pulse duration and integrated charge (ΔQ counts) recorded by the fast-reading current integrator under constant beam current. The plot shows data for 230 MeV FLASH proton beams delivered at the same nominal beam current on the CGMH SHI proton therapy machine. Even when the same pulse width is requested, both the actual delivered pulse duration and the total delivered charge exhibit noticeable variations during continuous beam output.

Figure 10. Measured integrated charge (ΔQ counts) versus actual pulse duration recorded by the fast-reading current integrator for 230 MeV FLASH proton beams delivered at different beam currents. The plot demonstrates that, even when the same pulse width is requested, both the actual delivered pulse duration and the total charge (dose) show noticeable variations during continuous beam output under varying beam current conditions.